\documentclass{jpsj3}

\usepackage{txfonts}

\title{SPARSE STATISTICAL MODELING IN CONDENSED MATTER PHYSICS}

\author{G. McGee$^1$ and S.V. Dordevic$^1$\thanks{dsasa@uakron.edu}}
\inst{$^1$The University of Akron, Akron, OH 44325, USA \\
} 

\abst{In this work we explore the possibility of using sparse statistical modeling 
in condensed matter physics. Sparse statistical modeling refers to techniques that 
focus on using a minimal number of predictors to explain a data set, and are particularly 
useful for high-dimensional data. The procedure is employed to two well known 
problems: elemental superconductors and heavy 
fermions, and was shown that in most cases performs better than 
other AI methods, such as machine or deep learning. More importantly, 
sparse modeling has two major advantages over other methods: the ability to 
deal with small data sets and in particular its interpretabilty.
Namely, sparse modeling can provide insight into the calculation process 
and allow the users to give physical interpretation of their results. 
We argue that many other problems in condensed matter physics would 
benefit from these properties of sparse statistical modeling.}

\begin{document}
\maketitle

\section{Introduction}
\label{introduction}
The last few years have see an avalanche of interest in applying AI methods,
in particular machine learning and deep learning, to problems in  
condensed matter physics \cite{ai-physics}. 
Thousands of papers have been published, addressing 
different tasks, with various levels of success. However, there are
two major problems with these calculations, which are often overlooked or ignored. 
First, available databases are usually very small by AI standards, 
with typically $\mathcal{O}(10^3-10^4)$ entries. In order to take
full advantage of AI calculations, input data sets should be several 
orders of magnitude larger \cite{google-dl}, i.e. $\mathcal{O}(10^6)$ or even larger. 
The second major problem with AI models in condensed matter physics is 
their non-transparency. They are typically performed
as "black box" calculation, where neither the authors nor the readers know
how or why certain results were obtained.

In this work we take a different approach. We employ the so-called sparse 
statistical modeling \cite{hastie-book,rish-book}, which seems to be ideally 
suited to address the two problems pointed out above. Namely, it has 
been known that sparse statistical modeling can be used and it works well with small
data sets. Sparse statistical modeling is also known for its interpretability, 
i.e. its ability to provide access into calculation process \cite{kwang25}. 
Here we apply sparse
modeling to two well known problems in condensed matter physics: elemental 
superconductors and heavy fermions. We show that, indeed, sparse modeling can
be successfully employed to both problems, and that usually 
performs better than traditional methods such as machine and deep learning.

The paper is organized as follows: in section \ref{methodology} we first make 
a brief introduction to the methodology used, i.e. the mathematical formulation 
of sparse statistical modeling. We then apply it to elemental superconductors
in section \ref{elemental} and to heavy fermion metals
in section \ref{hf}. Both problems have been studied before using other 
AI methods (machine and/or deep learning), 
which will allow us to directly compare and contrast the 
results. For both problems the databases used are very small (on 
the order of  $\mathcal{O}(10^2)$ entries), but with sparse modeling,
statistically significant results have been obtained, which in most
cases exceed those obtained with machine or deep learning.  
Finally, in section \ref{summary} we summarize the most important results and give 
a very positive outlook for the future use of sparse modeling in condensed matter 
physics.

\section{Methodology}
\label{methodology}

Sparse statistical modeling is based on the philosophical principle of parsimony,
also know as the Occam's razor \cite{rish-book}. In its simplest form, this 
principle postulates that the simplest solution is usually the most likely one. 
When applied to practical problems, it means that of all possible predictive variables,
only a small number is relevant for the response variable and is sufficient 
for learning an accurate predictive model. 

Mathematically, sparse statistical modeling can be formulated as a 
constrained optimization problem. For a real, noisy problem, the optimization
can we written as \cite{rish-book}:

\begin{equation}
\min_{\mathbf{x}} \|\mathbf{x}\|_0 \quad \text{subject to} \quad \|\mathbf{y} - \mathbf{A}\mathbf{x}\|_2 \leq \epsilon.
\label{sparse}
\end{equation}
where {\bf x} is the solution of the problem and {\bf y} is response variable. 
$\|\mathbf{x}\|_0$ is the pseudo-norm of the solution vector, and simply represents 
the number of non-zero elements of vector {\bf x}.  
Matrix {\bf A} is the data matrix, where the i$^{th}$ column represents the i$^{th}$
predictor variable and the j$^{th}$ row is the j$^{th}$ training data point.  
Solving this optimization problem results in a sparse solution for {\bf x} 
i.e. a solution that has a large number of elements equal to zero. The solution 
(minimization) process
selects those predictors that have the larges effect on the solution.  

There are numerous software packages that implement various routines for 
sparse modeling. We have adopted the one called SpaSM \cite{spasm}. 
This package has numerous routines for both supervised (classification
and regression) and unsupervised sparse learning. In particular, 
for classification the so-called Sparse Linear Discriminant 
Analysis (SLDA) is available, whereas for regression one can choose between 
Forward Selection, Least Angle Regression, LASSO or Elastic Net algorithms.

\section{Elemental Superconductors}
\label{elemental}

In this section sparse statistical modeling is employed to elemental 
superconductors. This is a well known problem in condensed matter physics 
and has been studied before using different theoretical, computational and
statistical approaches \cite{hirsch97}. The goal is to predict whether a material 
(in this case, an element) is a superconductor or not, based 
on its normal state (usually room temperature) properties. Additionally,
for those materials that are superconducting, the goal is also to predict 
their critical temperature T$_c$. This is a long standing problem in 
condensed matter physics, which is still waiting for a reliable solution.

For the purpose of these calculations we compiled a database of 27 physical 
and chemical properties (mostly at room temperature) of the first 100 elements 
in the periodic table. They are listed in Table~\ref{tab:properties} 
and include parameters which are known (or believed) to correlated 
closely with superconductivity, such as  electron-phonon coupling constant 
($\lambda$), electron density of states at the Fermi level (DOS(E$_F$)) or
Debye temperature ($\Theta_D$). The database also includes parameters for 
which it is currently not know how they affect superconductivity, such as
the work function (W), electrical conductivity ($\sigma$), 
electronegativity, electron affinity, and others. 
Our database more than doubles the previously used one, both in the number 
of elements, as well as the number of predictors \cite{hirsch97}. 
The properties were collected from various sources, such as 
Ref.~\cite{springer-handbook,elementdata}, with additional entries from 
Ref.~\cite{kittel-book,ashcroft-book,poole-book,poole-handbook}. 

We point out that, unlike previous studies, we studied 
elements that are superconducting both at ambient pressure, as well 
as at high applied pressures. Their superconducting temperatures are recorded 
in the database
under both conditions \cite{hirsch15,hamlin15,shimizu15,buzea05}, 
which allows us to conduct a complete and comparative study. Additionally, our 
study goes beyond simple statistical correlations, and allows the
models to learn not only how certain normal state properties {\it individually} 
affect superconductivity, but also how their {\it combinations} \cite{hirsch97}
affect superconductivity.


\begin{table}
    \centering
        \caption{Physical and chemical properties of elements used as predictors 
        of superconductivity. They are listed in random order, along with their commonly 
        used symbols and units.}
    \begin{tabular}{|l|l|c|c|}
    \hline
     &    Property &  Symbol  & Units \\
        \hline
  1   &    Fermi energy & E$_F$ & eV \\
  2   &    Hall coefficent & R$_H$  &  m$^3$/C       \\
  3   &    Electron-phonon coupling constant  & $\lambda$ &  \\
  4   &	   Work function   & W  &   eV     \\
  5 	  &	   Debye temperature  &  $\theta_D$  &  K     \\
  6   &    Electrical conductivity  & $\sigma$ &  $\Omega^{-1} m^{-1}$  \\ 
  7   &	   Electron affinity   & E$_{ea}$  &    eV     \\         
  8   &    Electronegativity  &  $\chi$ &         \\
  9   &    Volume magnetic susceptibility   &  $\chi_{v}$  &   \\
 10 	  &	   Specific heat   & C$_p$  &   J/(kg K)       \\
 11   &	   Crystal structure    &   &          \\
 12   &    Thermal conductivity   & $\kappa$  &   W/(K m)    \\
 13   &    Valence    &   &      \\
 14   &    Atomic mass     & A  &  u     \\
 15   &    Density of states & DOS(E$_F$) & states/eV \\   
 16   &    Density    &  $\rho$  &  kg/m$^3$       \\ 
 17   &    Ionization energy &  E$_{ion}$ & eV   \\
 18   &    Atomic radius   &   r  &   pm     \\
 19   &    Covalent radius   &  r$_{cov}$  &  pm     \\
 20   &    Melting temperature & T$_m$ & K \\
 21   &    Logarithmic average of phonon frequency  & $\omega_{log}$ &  K  \\
 22   &    Speed of sound & c  &  m/s       \\
 23   &    Bulk modulus   &  B   &   GPa     \\
 24   &    Young's modulus   &  E  &  GPa     \\
 25 	  &	   Thermal expansion coefficient  &  $\alpha$  &  1/K     \\
 26   &    Effective U  & U$_{eff}$ &  eV  \\ 
 27   &    Cohesive energy  & E$_{coh}$ &  eV/atom  \\ 
         \hline
    \end{tabular}   
    \label{tab:properties}
\end{table}

\subsection{Correlations}

Before attempting sparse calculations, it is instructive to check 
for any possible correlations between the parameters in the database. 
Namely, sparse modeling is a linear technique (Eq.~\ref{sparse}) 
and those correlation can have a detrimental effect on the model.
For that purpose, we have calculated Spearman, Pearson and Kendall correlation 
coefficients \cite{statistics-book} between all 27 predictors 
from Table~\ref{tab:properties}.

In Table~\ref{tab:correlations} we list twelve pairs of predictors which 
display statistically significant correlations, larger than approximately
0.75. We notice that seven are positive, while five are negative. They
have been discussed in the literature, for example the Wiedemann–-Franz law between 
electrical and thermal conductivity \cite{kittel-book,ashcroft-book}, or 
the relation between melting temperature and bulk modulus \cite{meyers-book}.
 For regression calculations, properties with positive correlations are 
removed (one from each pair) from the database, as it is unnecessary to 
include both of them in the calculations. In particular, we removed 
the speed of sound, electrical conductivity, atomic mass, atomic radius,
covalent radius, melting temperature, bulk modulus and the coefficient
of thermal expansion. On the other hand, for classification 
calculations all 27 predictors were used. It was found that removing 
predictors did not lead to any improvements in model performance.


\begin{table}
    \centering
        \caption{Statistically significant correlation between predictors from 
        Table~\ref{tab:properties}. Only those with coefficients greater than about 0.75 are 
        listed. Numbers in the brackets refer to Table.~\ref{tab:properties}.}
    \begin{tabular}{|l|l|c|c|c|c|}
    \hline
        Predictor 1 &  Predictor 2 & Spearman  &  Pearson & Kendall &  References \\
        \hline
       (5) Debye temp  & (22) speed of sound & 0.84  &  0.93  &     &  \cite{kittel-book,ashcroft-book} \\
       (6) el conductivity  & (12) th conductivity & 0.90  &  0.96  &  0.83  & \cite{kittel-book,ashcroft-book} \\
        (10) spec heat  & (14) atomic mass & -0.96  &    &  -0.88   &  \cite{kittel-book,ashcroft-book}\\
        (17) ionization en  & (18) atomic radius & -0.81  &  -0.82  &    &  \cite{petrucci-book} \\
        (17) ionization en  & (19) covalent radius & -0.84  &  -0.81  &    &  \cite{petrucci-book} \\
        (18) atomic radius  & (19) covalent radius & 0.90  &  0.92  &  0.76  &  \cite{kittel-book} \\
        (20) melting temp  & (23) bulk modulus & 0.82  &  0.75  &    &  \cite{meyers-book}   \\
        (20) melting temp  & (24) Young's modulus & 0.80  &    &    &   \cite{meyers-book}   \\
        (20) melting temp  & (25) th expan coef & -0.88  &    &  -0.73  & \cite{kittel-book,ashcroft-book} \\
        (20) melting temp  & (27) cohesive energy & 0.83  &  0.88  &    &  \cite{kittel-book,ashcroft-book} \\
        (23) bulk modulus  & (24) Young's modulus & 0.88  &  0.90  &  0.74  & \cite{goodno-book}  \\
        (25) th expan coef  & (27) cohesive energy & -0.82  &    &    & \cite{samsonov-handbook} \\
         \hline
    \end{tabular}   
    \label{tab:correlations}
\end{table}

\subsection{Classification}

In this subsection we discuss classification of the elements into different classes or groups. 
The idea is to construct a model capable of predicting 
whether an element (on more generally, a compound) is a superconductor or not,
based on its normal state properties.
For the purpose of these calculations, the first 100 elements in the periodic
table were divided into 4 classes (Table~\ref{tab:groups}).
Class 1 contains elements that are not superconductive under any conditions, such 
as Na, K, Cu, Au, etc. Class 2 contains elements that are not superconducting 
at atmospheric pressure (p=0), but become superconducting under pressure (p$>$0), 
such as B, Sr, Si, Ge, etc. Class 3 contains elements that are superconducting 
at atmospheric pressure, but application of pressure causes their critical 
temperature to either remain unchanged or decrease, such as Al, Zn, Mo, etc. 
Finally, class 4 has the elements which are superconducting at ambient pressure,
and application of pressure increases their critical temperature, such as La, V, Ti, etc. 

\begin{figure}
    \includegraphics[scale=0.5]{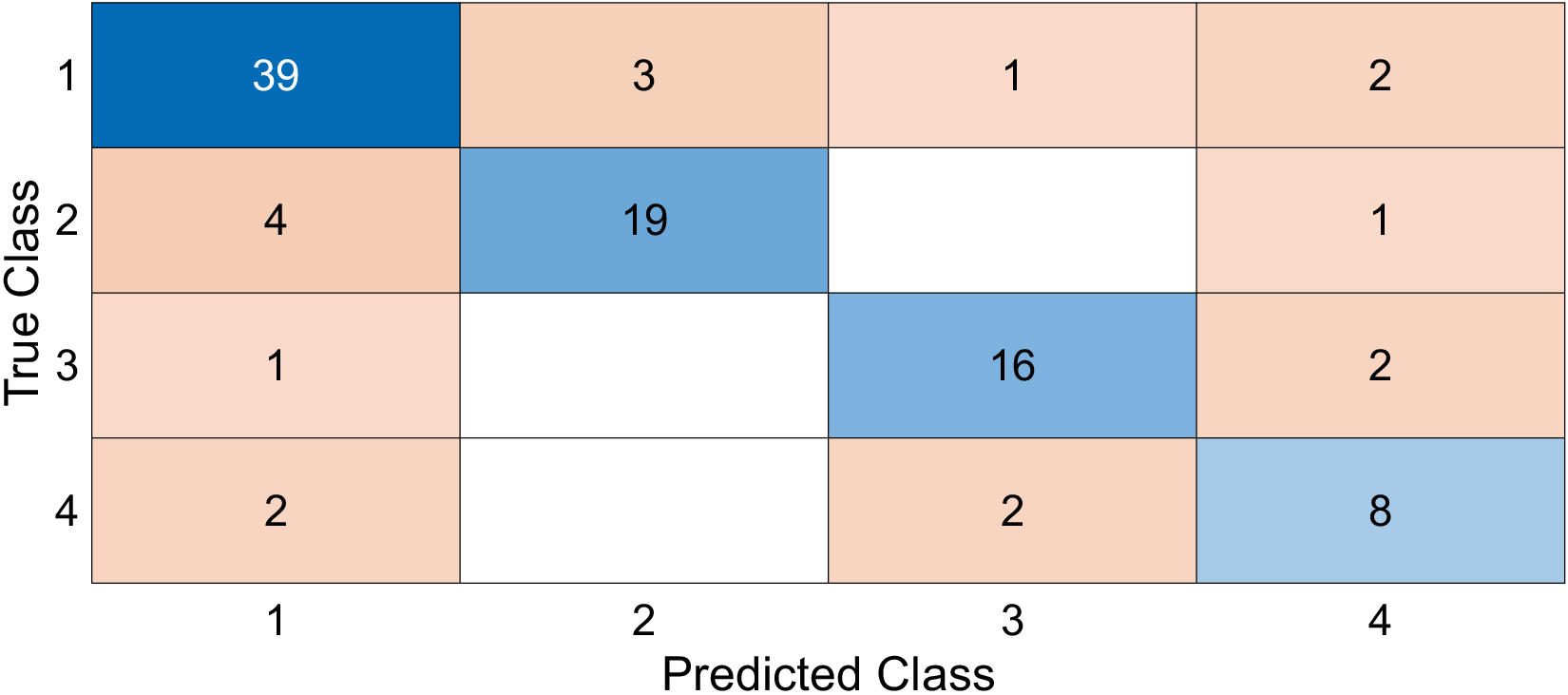}
    \caption{Confusion chart of classification with SLDA. 
    The four classes refer to Table~\ref{tab:groups}. 
    The sparse model achieved 82$\%$ prediction accuracy.}
    \label{fig:confusion}
\end{figure}

SLDA was performed on the whole database, i.e. all 27 predictors from Table~\ref{tab:properties}. 
It was found that removing some of the predictors, in particular those that are 
correlated, did not lead to any improvements in model performance. Before 
calculations were performed, all predictors were normalized to 0 mean value 
and standard deviation of 1. In Fig.~\ref{fig:confusion} we display the
confusion chart which is usually used to graphically represent both the accuracy
and errors in classification calculations. As can be inferred from the graph, 
the overall accuracy of the model is 82$\%$.


\begin{table}
    \centering
        \caption{Four classes of elements discussed in the text, on which 
        classification calculations were performed.}
    \begin{tabular}{|c|c|c|c|c|c|}
    \hline
        Class &  p=0  & p$>$0  &  T$_c$ under pressure & Examples  & Number of elements \\
        \hline
        1  & $\times$ & $\times$  &    &  Na, K, Cu, Au  &  45 \\
        2  & $\times$ & $\checkmark$   &   & B, Sr, Si, Ge   & 24   \\
        3  & $\checkmark$ & $\checkmark$   & T$_c$ does not increase   & Al, Zn, Mo  & 19    \\
        4  & $\checkmark$ & $\checkmark$    & T$_c$ increases       &    La, V, Ti   & 12  \\
         \hline
    \end{tabular}   
    \label{tab:groups}
\end{table}

\subsection{Regression}

Predicting critical temperatures of superconductors, including elemental superconductors,
is an old problem and numerous studies have approached it from various angles. 
In this work we used sparse modeling to address the problem. 
In spite of a very small data sets (only 31 superconducting elements at ambient pressure,
and 55 at high applied pressures), statistically significant results are achieved. 
More importantly, sparse modeling can provide us with access to the predictors 
that are most relevant for superconductivity. 

For the purpose of these calculations, only superconducting elements were used. 
All four sparse models mentioned above were 
tested, and the best results were achieved with Forward Selection. 
The model identified predictors (from Table~\ref{tab:properties}) which 
had the most significant predictive power for the critical temperature. 
The calculations were performed separately for superconductors at ambient pressure,
and for superconductors at high applied pressures. 
 
\begin{figure}
    \centering
        \vspace{0pt}
    \includegraphics[scale=0.5]{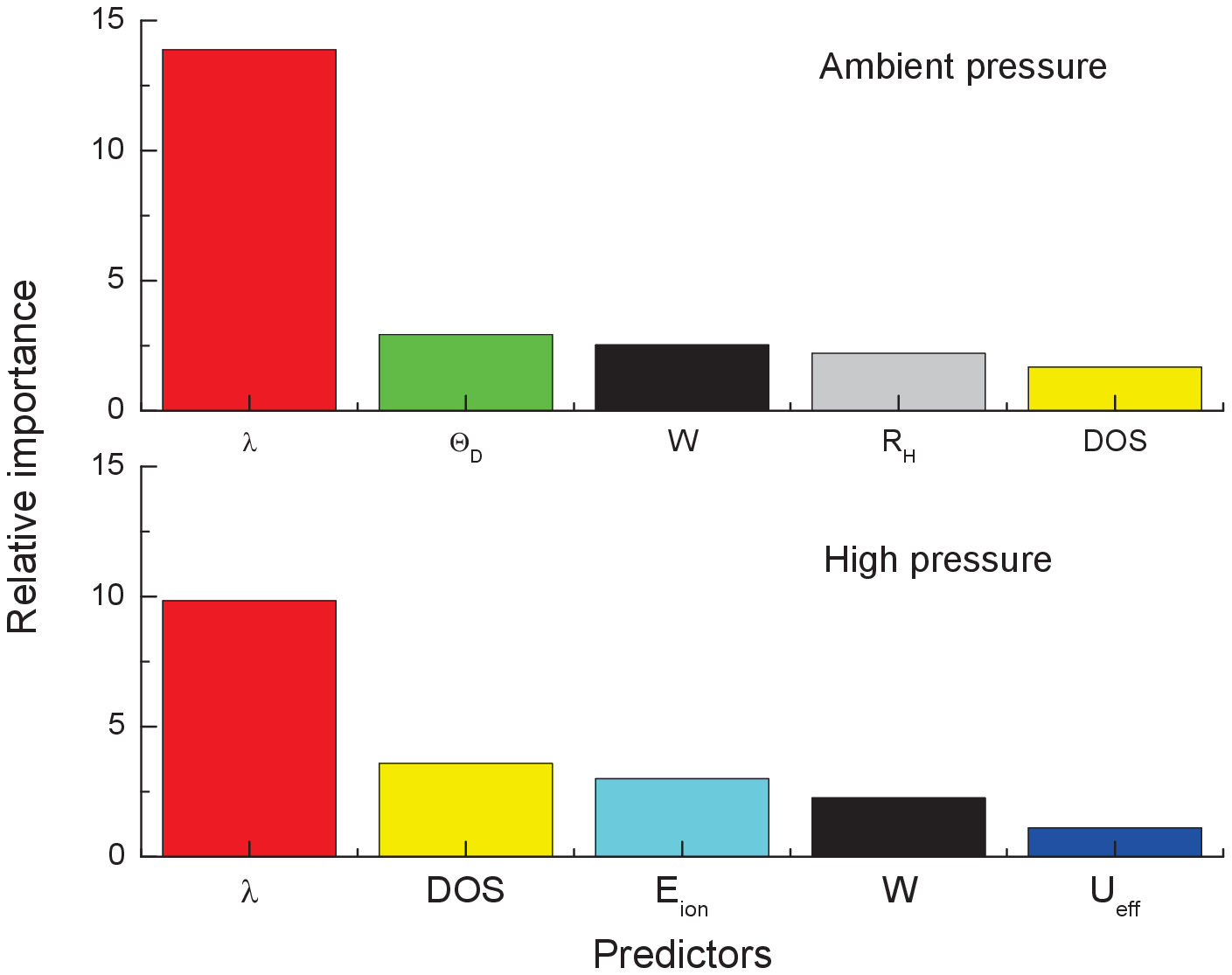}
        \vspace{0pt}
    \caption{(a) Most important predictors of superconductivity in elements at atmospheric pressure.
    (b) Most important predictors of superconductivity at high pressures. The symbols refer to 
    predictors from Table~\ref{tab:properties}. Note that vertical axes does not
    have any physical meaning; only the relative values among different predictors are important.}
    \label{fig:regression}
\end{figure}

In Fig.~\ref{fig:regression} we display 5 most important predictors for superconductivity, 
at both atmospheric pressure, as well as at high pressure. In both 
cases the most important parameter is the electron-phonon coupling constant 
($\lambda$). This supports the notion of electron-phonon mechanism of superconductivity in (most) 
elements. At atmospheric pressure, $\lambda$ is followed by Debye temperature, work function, 
Hall coefficient and DOS, in order of decreasing importance. At high pressures, $\lambda$ is followed
by DOS, ionization energy, work function and effective U. Perhaps somewhat surprising,
the work function (W) is identified as an important predictor in both cases. To the best
of our knowledge, its
effect on superconductivity in elements has not been systematically explored before. 

In order to assess the 
quality of the sparse model, we compare its predictions with other, physics-based
models: BCS, McMillan and SISSO formulas. In BCS theory \cite{bcs} the critical 
temperature is given as:

\begin{equation}
T_c = 1.14 \omega_D e^{-\frac{1}{\lambda}} 
\label{eq:bcs}
\end{equation}
where $\omega_D$ is the Debye frequency (or temperature) and $\lambda$ 
is the electron-phonon coupling constant
(predictors 5 and 3 in Table~\ref{tab:properties}).
In McMillan's theory \cite{mcmillan68} the critical temperature is given as: 

\begin{equation}
T_c = \frac{\omega_D}{1.45} exp{\Big(-\frac{1.04 (1+\lambda)}{\lambda - \mu^{*} 1.62 (1+\lambda)}\Big)}
\label{eq:mm}
\end{equation}
where $\mu^{*}$ is the so-called Coulomb pseudopotential, and its value is 
taken as 0.1 for all elemental superconductors. 
Finally, in the the so-called SISSO formula obtained with symbolic machine learning \cite{sisso}, 
the critical temperature is given as:

\begin{equation}
T_c = 0.09525 \frac{\lambda^4 \omega_{log}}{\lambda^3 + \sqrt{\mu^{*}}}
\label{eq:sisso}
\end{equation}
where $\omega_{log}$ is the logarithmic average of phonon frequency
(predictor 21 in Table~\ref{tab:properties}).

In Fig.~\ref{eq:bcs} we display values of T$_c$ predicted with all
three formulas. The size of the symbols used approximately correspond 
to the error bars of calculations.
As can be seen, BCS formula systematically overestimates 
the critical temperature, typically by an order of magnitude or more. 
McMillan and SISSO provide much better estimates. Most importantly,
sparse model can provide equally good estimate of T$_c$. We also notice from the figure 
that McMillan formula provides the best estimate for the critical temperature 
of elements with the lowest critical temperature (T$_c$ $\textless$ 1 mK).

\begin{figure}
    \centering
        \vspace{0pt}
    \includegraphics[scale=0.5]{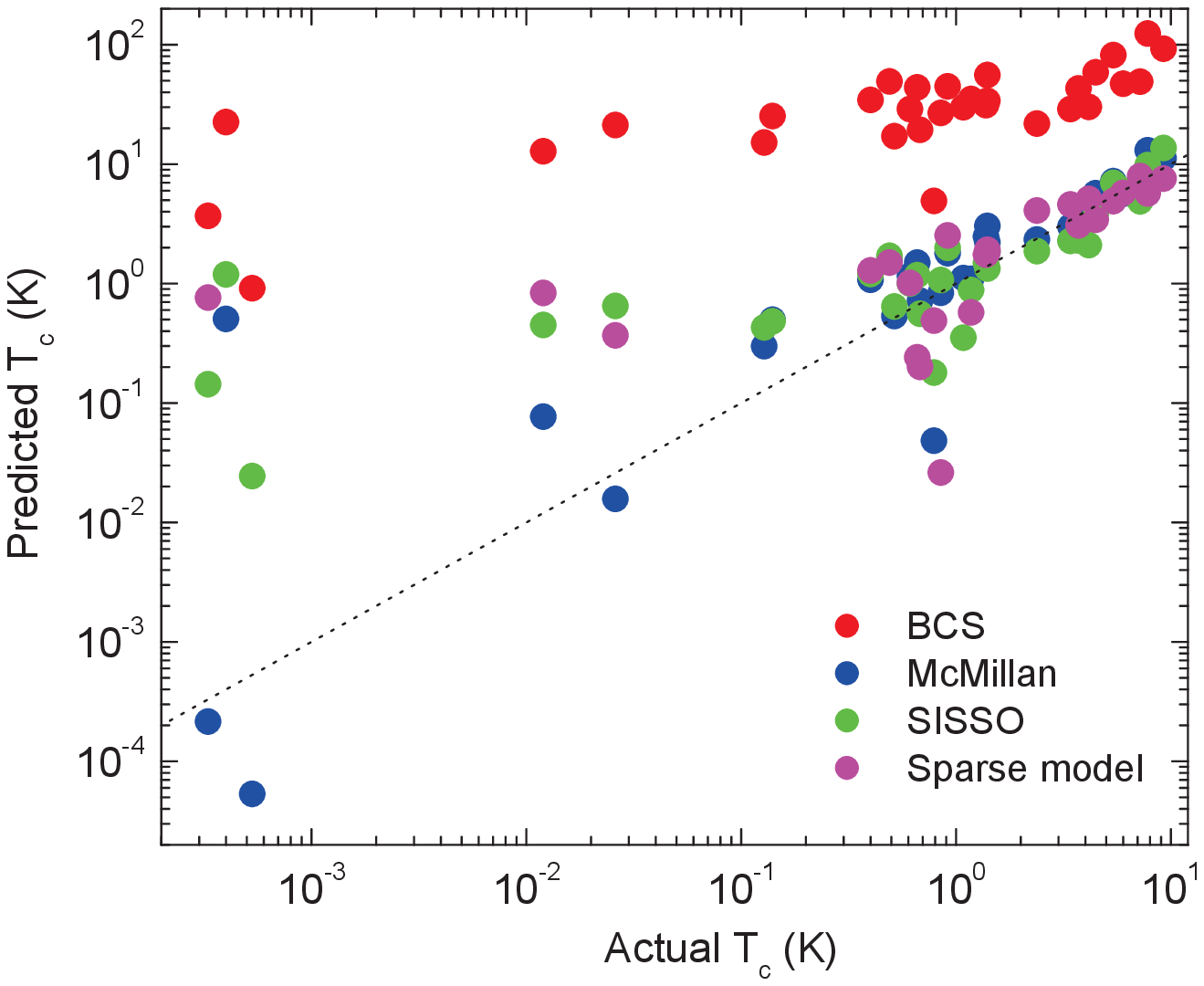}
        \vspace{0pt}
    \caption{Predicted critical temperature versus actual critical temperature 
    T$_c$ for elemental superconductors. The plot includes predictions from  
    BCS \cite{bcs}, McMillan \cite{mcmillan68} and  
    SISSO formulas \cite{sisso} discussed in the text, as well as 
    the predictions from the sparse model.}
    \label{fig:bcs}
\end{figure}

\section{Heavy Fermions}
\label{hf}

In this section we discuss the results of applying sparse statistical modeling
on another problem that was recently addressed using deep learning: predicting 
the properties of heavy fermion metals \cite{dlhf}. These materials 
usually contain lanthanide (for example Ce or Yb) or actinide (for example U or Pu) elements, 
which results in their unusual physical properties. In particular, the effective 
mass of charge carriers can be significantly enhanced compared to ordinary metals. 
In order to facilitate a direct comparison, we used the same database as in our previous study 
with deep learning \cite{dlhf}. The database contains 206 entries and includes
chemical compositions, coherence temperature, Sommerfeld coefficient, effective mass,
superconducting critical temperature and Neel temperature. 

\begin{figure}
    \centering
    \includegraphics[scale=0.4]{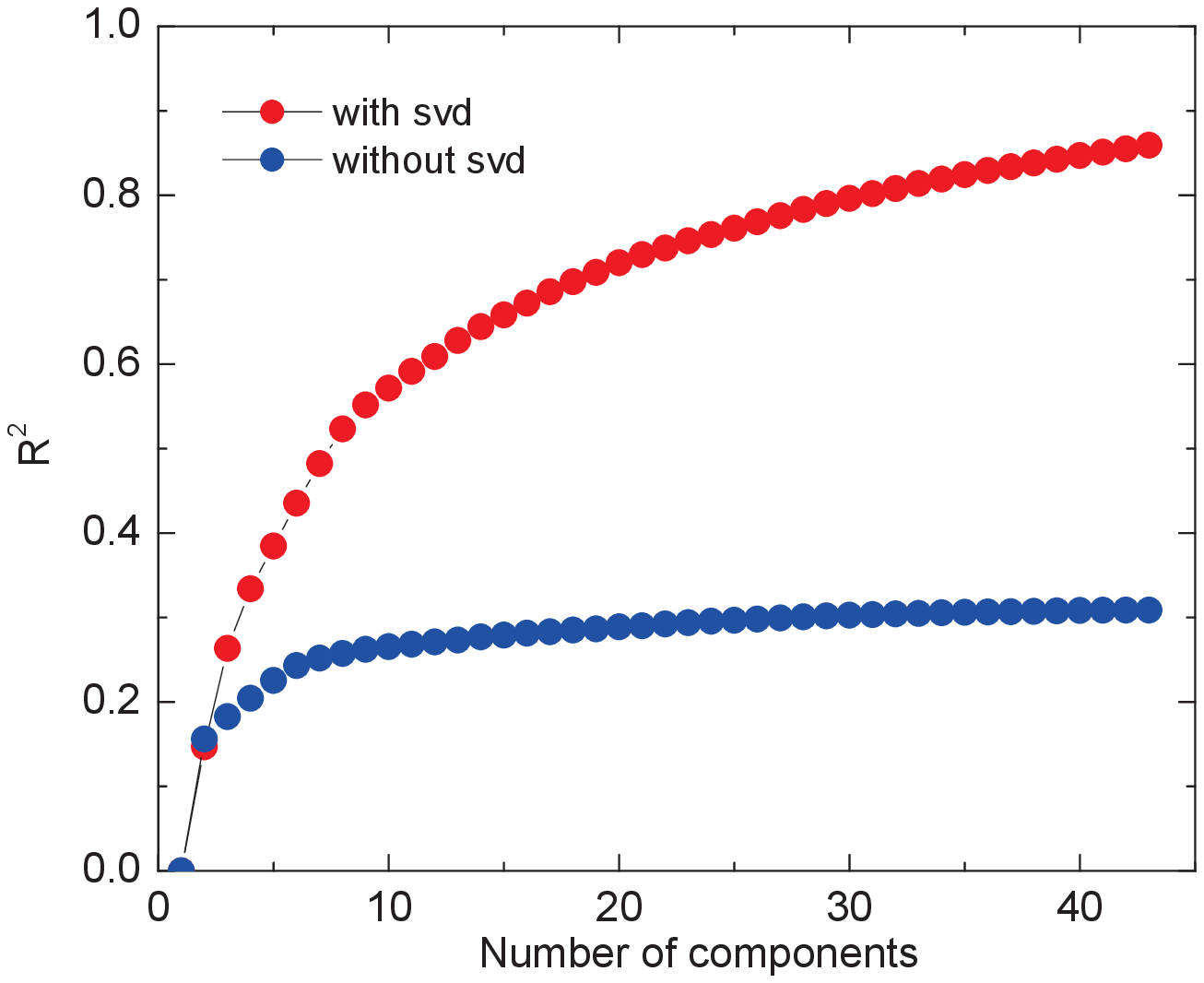}
    \caption{Statistical coefficient R$^2$ for a sparse model designed to predict 
    the effective mass of heavy fermions. Singular value decomposition is shown
    to dramatically increase the value of R$^2$.}
    \label{fig:train}
\end{figure}

Sparse regression models are used to learn the coherence temperature, Sommerfeld coefficient 
and effective mass. All compounds in the database contain 43 different elements, and only their
stoichiometric information was used  
as predictors. As can be seen from Fig.~\ref{fig:train}, the value of R$^2$ parameter \cite{statistics-book}
obtained for the effective mass
quickly saturates to approximately 0.3. However, when singular value decomposition \cite{svd}
is employed before sparse modeling, significant improvement is achieved and maximum
value of 0.86 attained. This value is higher than the one obtained using
deep learning (0.78) \cite{dlhf}. Classification calculations were also performed using SLDA and 
values of 0.92 and 0.79 were obtained for superconducting and anti-ferromagnetic 
heavy fermions, compared with 0.86 and 0.88 from deep learning \cite{dlhf}. 
We do acknowledge, however, that while SVD drastically improves predictive accuracy of the model
by handling collinearity and dimensionality, it compromises the direct physical interpretability 
of the sparse features.

\section{Summary}
\label{summary}
Sparse statistical modeling was shown to hold great potential for applications
in condensed matter physics. Its two main advantages are the ability to deal with (very) small
data sets and its interpretability. In the two examples presented, sparse modeling performed
comparable, and in many cases
better than deep learning and machine learning. For the superconducting properties of elements,
sparse models were able to provide us with an insight that was otherwise not available using 
other AI procedures. For heavy fermions, we were able to obtain statistical parameters that 
in most cases exceed those obtained deep learning. 
Interpretability of sparse models can be leveraged for active learning in order to accelerate 
the discovery of new superconductors and heavy fermions, buy targeting materials with certain
desired features identified by sparse models.
The two examples discussed here clearly demonstrate great
potential of sparse modeling for solving problems in condensed matter physics. There are 
numerous problems for which the available datasets are small, and which would benefit 
from the interpretability of sparse statistical modeling.

The databases used in these calculations are available upon reasonable request.






\end{document}